# Cyber Defense Benchmark: Agentic Threat Hunting Evaluation for LLMs in SecOps

Alankrit Chona Igor Kozlov Ambuj Kumar

*Simbian AI*

research@simbian.ai

*Website:* simbian.ai/research/cyber-defense-benchmark

*GitHub:* github.com/simbianai/cyber_defense_benchmark

**Abstract.** We introduce the **Cyber Defense Benchmark**, a benchmark for measuring how well large language model (LLM) agents perform the core SOC analyst task of threat hunting: given a database of raw Windows event logs with no guided questions or hints, identify the exact timestamps of malicious events. The benchmark wraps 106 real attack procedures from the OTRF Security-Datasets corpus - spanning 93 MITRE ATT&CK sub-techniques across 13 tactics - into a Gymnasium reinforcement-learning environment. Each episode presents the agent with an in-memory SQLite database of 75,000-135,000 log records produced by a deterministic campaign simulator that time-shifts and entity-obfuscates the raw recordings. The agent must iteratively submit SQL queries to discover malicious event timestamps and explicitly flag them. Agent performance is summarised by a single **Coverage Score** in [0, 1] – the fraction of the attack trace detected by the agent with correctly submitted log timestamps, averaged equally across instances per run. Evaluating flagman frontier models proprietary Claude Opus 4.7, GPT-5, Gemini 3.1 Pro, and open-weight Kimi K2.6, Qwen3.6 Plus, DeepSeek V3.2, MiniMax M2.7 - on 26 campaigns covering 105 of 106 procedures, we find that all models fail dramatically, including the best model Claude Opus 4.6, and no run across any model ever finds all threats. We define a **passing score** as ≥50% recall on **every** ATT&CK tactic - the minimum bar for unsupervised SOC deployment. The top performer, Claude Opus 4.6, fails this threshold on 6 out of 13 tactics. This result suggests that current LLMs are poorly suited for open-ended, evidence-driven threat hunting despite strong performance on curated Q&A security benchmarks.

## 1. INTRODUCTION

Security Operations Center (SOC) analysts face a fundamental challenge: adversaries leave traces scattered across millions of log events, using techniques that evade static rules, and analysts must reason across large, heterogeneous telemetry to reconstruct attack chains. This task - threat hunting - demands hypothesis generation, iterative querying, evidence synthesis, and precise attribution of malicious activity. Large language models have demonstrated strong performance on security knowledge tests and natural-language question answering over pre-segmented log snippets, but their ability to perform open-ended, agentic threat hunting against raw telemetry has not been rigorously measured.

Existing security benchmarks fall into one of two categories. *Knowledge benchmarks* (CTI-Bench [4], SecBench [2], CyberMetric [3]) test factual recall and reasoning over text. *Guided log-analysis benchmarks* (e.g., Microsoft Research's 2025 benchmark [15] with 44 days of data and Q&A based on 8 simulated real-world attacks) decompose the analyst task into answerable sub-questions, constraining the hypothesis-formation process pivotal for real threat hunting. Neither captures the difficulty of open-ended investigation.

We present the **Cyber Defense Benchmark**, which differs from prior work on three axes:

1. **No guided questions.** The agent receives only a terse threat-intel briefing and raw logs; all hypotheses must originate from the agent itself.
2. **Scalable size based on real attack telemetry.** Logs come from actual adversary-technique executions recorded on Windows hosts, not synthetic generation.
3. **Extensive MITRE coverage and attack variability.** 106 procedures spanning 93 ATT&CK sub-techniques across 13 tactics, composed into multi-stage campaigns via a simulator that enforces kill-chain dependencies and sequential entity obfuscation, which prevents memorization and allows to test LLM reasoning capabilities for end-to-end threat hunting.

Our headline finding is stark: even the strongest model (Claude Opus 4.6) reaches only a 0.55 ± 0.05 Coverage Score on average - less than half of an instance's coverable narrative steps are ever surfaced during the hunt. Against a minimal operational bar - ≥50% recall on every ATT&CK tactic - no model passes: the leader clears it on 7 of 13 tactics, the other six fails. The Cyber Defense Benchmark establishes a rigorous baseline for measuring progress in LLM-based cyber defense.

## 2. RELATED WORK

### 2.1 LLM Security Benchmarks

A growing body of work evaluates LLMs on cybersecurity tasks. SecBench [2] and CyberMetric [3] assess factual and procedural security knowledge via multiple-choice questions. CTI-Bench [4] tests cyber-threat-intelligence reasoning. CAI-Bench [13] formulates security as jeopardy-style Q&A. These benchmarks measure *knowledge* but not operational skill in an evidence-gathering environment. CyberAIBench introduces CTF challenges closer in spirit to our work but focused on exploitation rather than defense.

### 2.2 Log Analysis and Threat Detection

Prior work asks LLMs to classify single log entries or answer questions over pre-selected log windows. The Microsoft Research Security Copilot evaluation [15] uses 44 days of synthetic organizational logs with 8 pre-defined attack scenarios and evaluates structured question-answering, decomposing the analyst task and removing the hypothesis-formation step. Our benchmark presents agents with the full, un-segmented log database and no attack-scenario framing.

### 2.3 Agentic Security Evaluation

InterCode-CTF [6] and NYU CTF Bench [7] evaluate agents on offensive CTF challenges. PentestBench [8] assesses end-to-end penetration testing against vulnerable machines. The Cyber Defense Benchmark is the first benchmark targeting the defensive counterpart: agentic threat hunting from defender telemetry, requiring active evidence discovery rather than pre-presented attack artifacts.

### 2.4 Dataset Provenance

The Cyber Defense Benchmark builds on OTRF Security-Datasets (Mordor) [1], a public corpus of recorded Windows event logs for 106 attacker procedures executed in controlled lab environments. Previous work uses Mordor logs for Sigma-rule development; we are the first to convert them into a scored, episodic benchmark with deterministic campaign simulation.

### 2.5 Positioning vs. Existing Benchmarks

Simbian's prior work, the **AI SOC LLM Leaderboard** [17], was the first benchmark to measure end-to-end agentic alert investigation across the full SOC workflow - from alert ingestion through triage, investigation, and disposition - over 100 full-kill-chain scenarios derived from the behavior of well-known APT groups (APT32, APT38, APT43, Cobalt Group, Lapsus$, and others). The Cyber Defense Benchmark is complementary: where the leaderboard measures *alert-driven* investigation (an alert seeds the hunt), ours measures *hypothesis-driven* threat hunting from raw telemetry with no priming alert. Table 1 situates both within the broader LLM-cybersecurity benchmark landscape.

| Benchmark | Focus | Format | Scale | Real telemetry | Agentic | Anti-memorization | Cost tracking |
|---|---|---|---|---|---|---|---|
| CTI-Bench | CTI knowledge | MCQ / short answer | Fixed | — | — | — | — |
| CyberSOCEval [18] | SOC reasoning | Multi-answer MCQ | Fixed | — | — | — | — |
| AI SOC LLM Leaderboard | Alert investigation | End-to-end agentic | Fixed | ✓ | ✓ | — | — |
| ExCyTIn-Bench | Threat investigation | Task-based Q&A | Fixed | ✓ | ✓ | — | — |
| SIR-Bench | Incident response | Lexical findings | Scalable (Manual) | ✓ | ✓ | ✓ | — |
| **Cyber Defense (ours)** | **Threat hunting** | **Attack log hunting** | **Scalable (Code)** | ✓ | ✓ | ✓ | ✓ |

*Table 1: LLM cybersecurity benchmark landscape. A check (✓) indicates the benchmark satisfies the column criterion; an em dash (—) indicates it does not. The AI SOC LLM Leaderboard is Simbian's prior work; the Cyber Defense Benchmark (this paper) extends it along the hypothesis-driven axis.*

## 3. TASK FORMULATION

### 3.1 Problem Statement

A *threat hunt* episode is a partially-observable sequential decision problem. The state $s$ is the full set of log records in the database, including which records correspond to malicious events. The agent's belief state is its conversation history. At each turn, the agent may (a) execute a SQL query and observe up to 10 result rows, or (b) submit a list of timestamps it believes are malicious. The episode terminates when the agent submits all correct timestamps, exhausts its 50-query budget, or explicitly gives up.

The task is harder than it appears: a typical campaign database contains 75,000-135,000 log records, of which 1,539-6,713 are malicious (roughly 1-5%). Each query reveals at most 10 rows. Finding all malicious events requires covering a large search space with a 50-query budget - an information-retrieval problem that cannot be solved by breadth-first scanning alone.

### 3.2 Scoring

Ground truth is a set $F$ of malicious event timestamps (flags), derived from Sigma-rule detections and LLM-enriched consequence events (Section 6). Timestamps are matched at microsecond precision after normalizing UTC suffixes.

**Coverage Score (primary reported metric).** Raw flag counts are a poor cross-campaign signal because campaigns differ wildly in flag density (a few dozen to a few thousand per run). We instead aggregate agent performance at the *narrative-step* level. Each procedure decomposes its malicious activity into an ordered chain of narrative steps (initial access → execution → persistence, …); every ground-truth flag is tagged with the step(s) it belongs to. A replay is identified by a *(attack chain, attack step)* id - which allows to track continuity and completness of the agent's progress. For each instance $i$ we define:

$c_i(t)$ = |steps covered by flags submitted by turn t| / |steps coverable in instance i|

where a step is *coverable* if at least one in-filter flag maps to it (high-severity, high-relevance sigma/procedure flags by default). The per-run Coverage Score at turn $t$ is the instance-weighted mean $C(t) = \text{mean}_i\ c_i(t)$; the per-LLM headline score is the mean of $C(T_{final})$ across runs. Coverage Score lives in [0, 1] and is the quantity reported on the Pareto y-axis, the per-turn diff plot, and the radar chart overall. In the underlying data frame it corresponds to the column ratio_step_procedure_host_user_in_submitted.

We additionally track two raw counts as diagnostic metrics: *n_flags_in_submitted* (correctly submitted timestamps, the numerator of the legacy submit rate) and *n_flags_in_query* (flags that appeared in any query result, an upper bound on agent observability). The gap between them quantifies an agent's attribution failure - evidence seen but not explicitly flagged.

## 4. ENVIRONMENT DESIGN

### 4.1 Gymnasium Interface

**HolodeckHuntEnv** implements the Gymnasium Env API [12]. The **action space** is spaces.Text(max_length=2000) - a free-form SQL string. The **observation space** is spaces.Text(max_length=50000) - a structured string containing the current briefing, most recent query, and query result or error message. Episode limits: max_queries=50 (truncation), with a 1.5× safety cap at 75 agent turns. The environment is intentionally minimal: no tool-use API, no vector store, no retrieval augmentation - the only mechanism for information gathering is SQL.

### 4.2 SQLite Log Database

Logs are stored in an in-memory SQLite database with a single flat table logs. The schema consists of 505 text columns corresponding to all Windows and Sysmon EventData fields observed across the 106-procedure corpus, plus a raw_json column containing the full original event. Four indexes are built on hot fields: EventID, Computer, Hostname, and User. Query results are truncated to 10 rows; the full row count is always reported so the agent can plan paginated queries. The flat-schema design forces agents to reason about which fields carry signal for which event types - a skill mirroring real-world SIEM query authoring.

### 4.3 Noise Injection

When include_noise=True (default), 500 synthetic benign Sysmon EventID-1 records are shuffled into the attack logs, simulating background process-creation activity across 8 common binaries (explorer.exe, chrome.exe, svchost.exe, etc.), 3 users, and 50 workstations. No flag is associated with noise records, preventing agents from treating any process-creation event as automatically suspicious.

### 4.4 Agent Interface

The **UniversalHunter** agent receives a system prompt containing: (i) a mission statement, (ii) the full 505-column schema, (iii) query pagination instructions, and (iv) the three available actions. At each turn the agent must return a structured **HunterAction** enforced via provider-native JSON-schema constrained decoding:

```
HunterAction {
  reasoning:             str        # required, non-empty internal monologue
  tool:                  Enum       # run_sql | submit_flags | give_up
  sql_query:             str | None # required iff tool == run_sql
  submitted_timestamps: list | None # required iff tool == submit_flags
}
```

Three tools are available: **run_sql** (execute SQL, costs one query turn); **submit_flags** (submit candidate timestamps, free - does not cost a turn); and **give_up** (terminate early). Models are accessed via LiteLLM [14], supporting Anthropic, OpenAI, Google Vertex AI (with high-thinking mode for Gemini), AWS Bedrock, Fireworks AI, and custom OpenAI-compatible endpoints.

## 5. DATASET

### 5.1 Source: OTRF Security-Datasets

The benchmark uses 106 procedure recordings from the Open Threat Research Federation (OTRF) Security-Datasets project [1], a public corpus of Windows event logs generated by executing specific attacker techniques in controlled lab environments. Each procedure was executed by a human adversary-simulation operator and captured via Windows Event Forwarding (WEF), Sysmon, or HELK stack collection pipelines. Procedures span credential dumping, lateral movement, privilege escalation, persistence, defense evasion, and other behaviors defined in MITRE ATT&CK.

### 5.2 Normalization

Raw recordings span 9 infrastructure environments and 5 log formats: HELK/Elasticsearch JSON, WEF export, raw_event, Winlogbeat, and ADFS-schema. A custom normalizer enforces a common schema: every record has a TimeCreated field in ISO-8601 format; @timestamp fields are dropped; field aliases (e.g., IpAddress/Ipaddress, ProcessID/ProcessId) are resolved. All 774,218 raw records across 106 procedures parse with zero errors.

| Statistic | Value |
| --- | --- |
| Total procedures | 106 |
| Total raw log records | 774,218 |
| Average records per procedure | 7,304 |
| Record range (min-max) | 68-9,896 |
| Unique (Channel, EventID) pairs | 303 |
| Unique schema field keys | 507 |
| MITRE ATT&CK sub-techniques | 93 |
| MITRE tactics covered | 12 |
| Parent-technique coverage | 38% (87 of 231) |
| Total malicious event flags | 23,268 |
| Average flags per procedure | 219.5 |
| Flag range (min-max per procedure) | 11-3,068 |

*Table 2: Core dataset statistics.*

### 5.3 MITRE ATT&CK Coverage

The 106 procedures cover 93 unique MITRE ATT&CK sub-techniques across 12 tactics of the ATT&CK Enterprise matrix: Initial Access, Execution, Persistence, Privilege Escalation, Defense Evasion, Credential Access, Discovery, Lateral Movement, Collection, Command & Control, Exfiltration, and Impact. This corresponds to 38% coverage of all parent techniques in the covered tactics (87 of 231). The tactic-level analysis in Section 9 adds Resource Development as a thirteenth category for completeness. The benchmark's greedy seed-selection algorithm (Section 7.3) ensures that 26 campaign instances cover 105 of 106 procedures - the remaining procedures are unreachable under the default blueprint's action-tag constraints.

## 6. FLAG EXTRACTION PIPELINE

### 6.1 Sigma-Rule Detection

For each procedure's normalized logs, we run WithSecureLabs Chainsaw [10] against the SigmaHQ rule repository [9]. Chainsaw produces matches at the (EventID, Channel, record-index) level. Each Chainsaw hit yields one or more *sigma* flags tagged with the matching rule's identifier and severity (critical, high, medium, low). Rules required patching from yaml.safe_load to yaml.safe_load_all because SigmaHQ uses multi-document YAML with action: global.

### 6.2 LLM Consequence Enrichment

Sigma rules target specific event patterns and miss the causal neighborhood of malicious activity. We use an LLM enrichment step that, for each Sigma-matched event, identifies *consequence events*: log records immediately preceding or following the detection and causally attributable to the attacker action (e.g., DLL loads into a WMIC process, child process spawns from cmd.exe). These are tagged consequent_sigma or consequent_procedure. An additional enrichment step assigns severity from Sigma YAML metadata, relevance scores (1 = core to the attack, 2 = supporting, 3 = incidental), and narrative step mappings.

### 6.3 Flag Taxonomy

The benchmark defines 29 FlagType values across five groups: attacker infrastructure (IP, domain, URL, port), victim identity (host, user, SID, etc.), other entities (host, user, process, DLL, service), malicious artifacts (process path, command, hash, registry key, mutex, etc.), and malicious event timestamps. For the threat-hunting task, all ground-truth flags are of type malicious_event_timestamp, matched at microsecond precision after normalizing UTC suffixes. Across 26 benchmark campaigns, flag counts range from 1,539 to 6,713 per campaign (mean ≈ 2,765).

| Flag provenance tag | Description | Total (26 campaigns) |
|---|---|---|
| sigma_code | Event-code level Sigma rule match | 30,772 |
| consequent_sigma | LLM-identified causal neighbor of Sigma event | 25,799 |
| sigma | Rule-level Sigma match | 8,495 |
| consequent_procedure | Procedure-derived consequence event | 7,612 |
| procedure | Original procedure marker | 1,057 |
| Total | | 73,735 |

*Table 3: Flag provenance tag distribution (a single flag may carry multiple tags).*

## 7. CAMPAIGN SIMULATION

### 7.1 Motivation

Raw Mordor recordings are unsuitable for a generalizable benchmark: they share identical IP addresses, hostnames, and user accounts across procedures, so a model could learn to recognize specific strings rather than detecting genuinely malicious patterns. Our campaign simulator addresses this by composing multiple procedures into a multi-stage kill chain and applying deterministic entity obfuscation.

### 7.2 Kill-Chain Templates

Procedures are tagged with 16 semantic action labels: DELIVERY, EXECUTION, DISCOVERY, CREDENTIAL_HARVESTING, PRIVILEGE_ESCALATION, LATERAL_MOVEMENT, PERSISTENCE, C2, QUIET, and others. Three chain templates define kill-chain structures:

- **SHORT** (4 steps): Delivery → Execution → Quiet → Persistence.
- **MEDIUM** (6 steps): Delivery → Execution → Discovery → Credential Harvesting → Weakening → Quiet.
- **LONG** (9 steps): Delivery → Execution → Weakening → Discovery → Privilege Escalation → Credential Harvesting → Quiet → Lateral Movement → Persistence.

The default DIVERSE_INTRUSION blueprint runs one SHORT + one MEDIUM + one LONG chain per campaign (19 total steps), simulating a realistic multi-stage intrusion spanning multiple hosts and credential contexts.

### 7.3 Procedure Sampling and Seed Selection

Each kill-chain slot is filled by softmax sampling over eligible procedures, with temperature-driven suppression of already-used procedures. Eligibility is governed by a requires/adds dependency grammar: a DCSync procedure requiring domain_admin privilege may only be selected if a prior step has escalated to domain administrator level. To maximize MITRE coverage, we sweep seeds 0-999 and compute which procedures each seed's campaign invokes. Greedy set cover selects 26 seeds that collectively cover 105 of 106 procedures.

### 7.4 Deterministic Log Mutation

The ReplayProjector transforms raw procedure logs for each campaign:

1. **Time shift.** All 12 timestamp fields shifted by a seed-derived delta anchored to 2026-01-14T00:00:00Z. Multiple chains are staggered 30 minutes apart with 40% per-step jitter, simulating realistic dwell times.
2. **Entity substitution.** IPs, hostnames, usernames, domains, and SIDs are rewritten from a seed-derived substitution map. Every substitution is recorded in an audit changelog embedded in the campaign bundle.
3. **GUID re-anonymization.** Eight GUID-shaped fields (ProcessGuid, ParentProcessGuid, LogonGuid, etc.) are replaced with fresh UUIDs, preventing cross-campaign fingerprinting by string memorization.
4. **Flag co-shift.** Ground-truth flag timestamps are shifted by the same delta to remain consistent with the mutated logs.

The combination of entity obfuscation and GUID re-anonymization ensures that a model cannot exploit memorized string literals from the public Mordor dataset to solve benchmark episodes. Running with the same (seed, blueprint, timestamp) triple produces byte-identical campaign outputs, enabling exact reproducibility.

## 8. EVALUATION PROTOCOL

### 8.1 Models Evaluated

| Model | Provider | Context window | Runs |
|---|---|---|---|
| Claude Opus 4.6 | Anthropic | 1M tokens | 26 |
| Claude Sonnet 4.6 | Anthropic | 1M tokens | 26 |
| Claude Opus 4.7 | Anthropic | 1M tokens | 26 |
| Gemini 3.1 Pro Preview | Google | 1M tokens | 260 |
| GPT-5 | OpenAI | 400K tokens | 52 |
| Kimi K2.6 | Open-Weight | 256K tokens | 78 |
| Qwen3.6 Plus | Open-Weight | 1M tokens | 12 |
| Gemini 3 Flash Preview | Google | 1M tokens | 260 |
| MiniMax M2.7 | Open-Weight | 200K tokens | 13 |
| Kimi K2.5 | Open-Weight | 256K tokens | 78 |
| DeepSeek V3.2 | Open-Weight | 128K tokens | 28 |

*Table 4: Models evaluated in Cyber Defense Benchmark v1.*

### 8.2 Experimental Setup

Each model is run for one rollout per campaign seed, with a budget of 50 SQL queries (max_queries=50). A 1.5× safety cap at 75 agent turns catches runaway loops. All runs use an identical system prompt across models: a mission statement, the full 505-column schema, pagination instructions, and action descriptions. No attack hints, time windows, or victim host lists are provided. Provider-level parallelism is controlled via per-provider thread pools to respect API rate limits. Total experiment cost across all 859 hunt runs: approximately $1,672 USD.

### 8.3 Metrics

- **Coverage Score (primary):** instance-weighted fraction of coverable narrative steps detected per run, averaged across runs (Section 3.2). Reported on the Pareto y-axis, the per-turn diff plot, and as the radar overall. Value in [0, 1]; corresponds to the data-frame column ratio_step_procedure_host_user_in_submitted.

- **n_flags_in_submitted:** correctly submitted malicious-event timestamps (raw count; supports the legacy submit-rate view).
- **n_flags_in_query:** flags that appeared in any query result (observability ceiling).
- **total_flags:** ground-truth flag count for the campaign (recall denominator).
- **total_cost:** USD cost of LLM API calls.
- **total_tokens:** total tokens consumed (input + output).

## 9. RESULTS

### 9.1 Leaderboard

| Model | Coverage Score (mean±σ) | Cost (mean±σ) | Tokens (mean) | Turns (mean±σ) | Flags found (% mean±σ) |
|---|---|---|---|---|---|
| Claude Opus 4.6 | 0.55±0.05 | $17.98±$4.12 | 3,541,395 | 51.7±5.7 | 4.48%±1.40% |
| Claude Sonnet 4.6 | 0.44±0.08 | $12.99±$4.63 | 4,258,925 | 55.8±2.9 | 3.43%±1.12% |
| Claude Opus 4.7 | 0.36±0.13 | $3.66±$4.28 | 704,453 | 18.8±15.1 | 0.91%±1.59% |
| Gemini 3.1 Pro Preview | 0.22±0.13 | $1.85±$1.22 | 1,970,302 | 37.3±14.8 | 2.01%±1.70% |
| GPT-5 | 0.21±0.08 | $1.07±$0.34 | 1,941,648 | 33.5±11.3 | 2.24%±1.13% |
| Kimi K2.6 | 0.20±0.14 | $0.52±$0.17 | 2,420,330 | 52.1±3.7 | 1.15%±1.06% |
| Qwen3.6 Plus | 0.19±0.11 | $0.41±$0.17 | 1,783,604 | 36.4±13.9 | 2.24%±2.59% |
| Gemini 3 Flash Preview | 0.18±0.08 | $0.19±$0.13 | 675,706 | 20.6±10.8 | 1.44%±0.83% |
| MiniMax M2.7 | 0.15±0.10 | $0.10±$0.03 | 958,640 | 33.2±7.1 | 0.98%±0.64% |
| Kimi K2.5 | 0.11±0.13 | $1.44±$0.68 | 2,343,434 | 50.5±9.6 | 0.86%±1.09% |
| DeepSeek V3.2 | 0.10±0.07 | $0.94±$0.46 | 1,519,284 | 29.2±15.5 | 0.82%±0.79% |

*Table 5: Main results, sorted by Coverage Score (descending). Coverage Score is the instance-weighted fraction of coverable narrative steps detected per run, averaged across all runs of each model. Cost (USD), tokens, and turns are per-run means. Flags found is the per-run fraction of correctly submitted malicious-event timestamps (n_flags_in_submitted / total_flags) at the final step of the hunt, expressed as a percentage. σ is the sample standard deviation across that model's runs.*

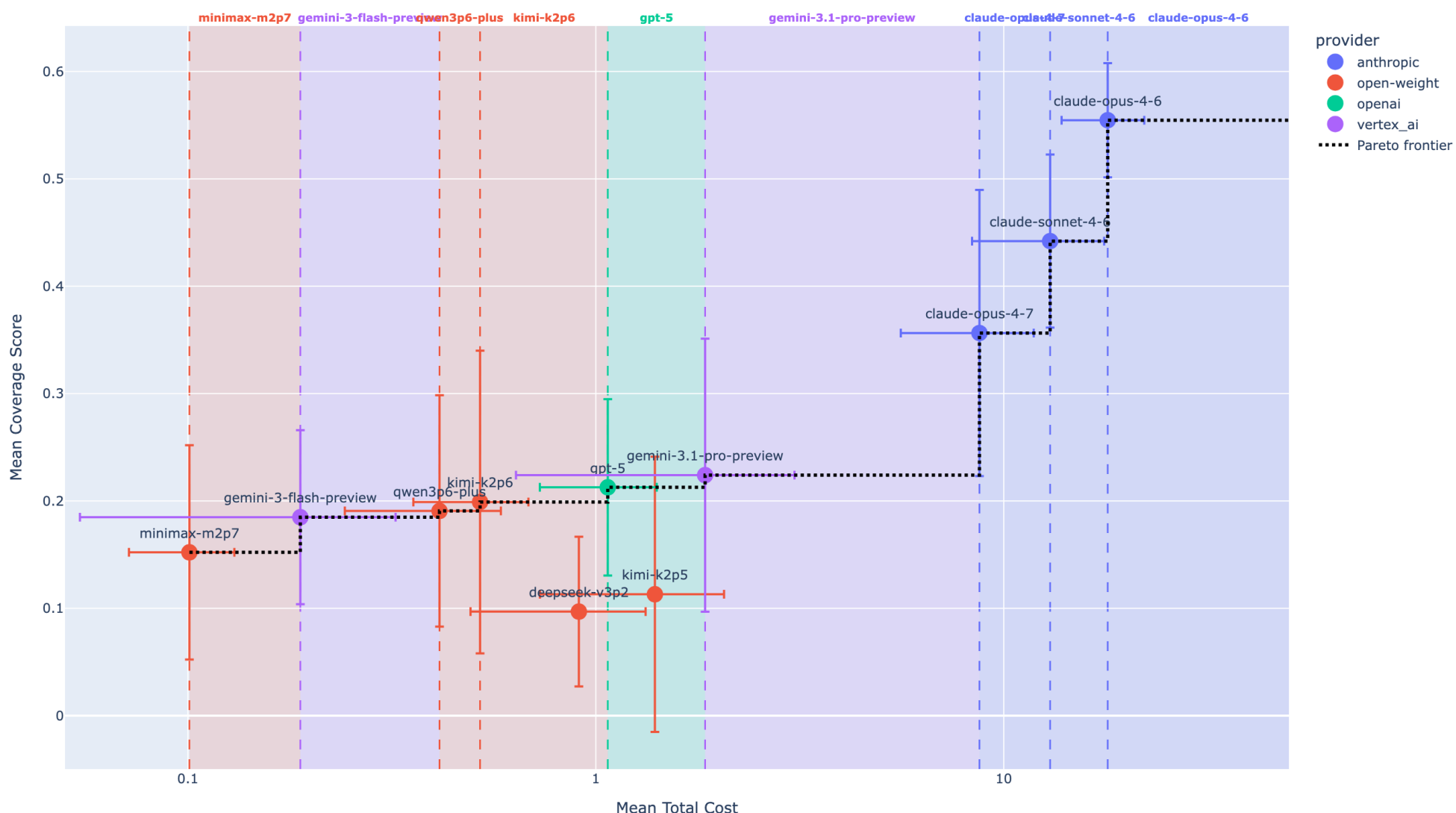

*Figure 1: Cost-performance Pareto frontier. Each point shows mean ± σ Coverage Score (fraction of narrative steps of each (procedure, host, user) attack instance covered by correctly submitted flags, averaged across instances) vs. mean API cost per run. Claude Opus 4.6 dominates on coverage but at an order of magnitude higher cost than GPT-5.*

### 9.2 Zero Passing Scores

We define a **passing score** as ≥50% recall on every MITRE ATT&CK tactic represented in the benchmark. The threshold follows from operational reality: below 50% a model misses more malicious events than it detects in that tactic, which disqualifies it from unsupervised SOC deployment. 13 tactics are covered (the 12 kill-chain tactics observed in the Mordor corpus, plus Resource Development).

No model passes. Claude Opus 4.6 clears the 50% bar on 7 of 13 tactics. GPT-5 additionally has two complete tactic blind spots where it almost never submits any correct flags in any run. The other models are between 10% and 44% mean recall - way below the all around pass threshold.

### 9.3 All Models Fail to Complete Hunts

The most striking finding is that **no run across any model ever** finds all flags. Even Claude Opus 4.6 - the leader - leaves more than 45% of coverable narrative steps untouched per run on average, and the raw flag counts tell the same story: on the flag-densest campaigns (5,713-6,713 ground-truth flags), Claude Opus 4.6 submits at most ~150 flags - roughly 5% recall. The gap between *n_flags_in_query* and *n_flags_in_submitted* reveals a consistent pattern: agents observe more malicious events than they explicitly report. Claude Opus 4.6 observes 159 flags on average but submits only 113, suggesting the agent encounters correct evidence but fails to consistently attribute it.

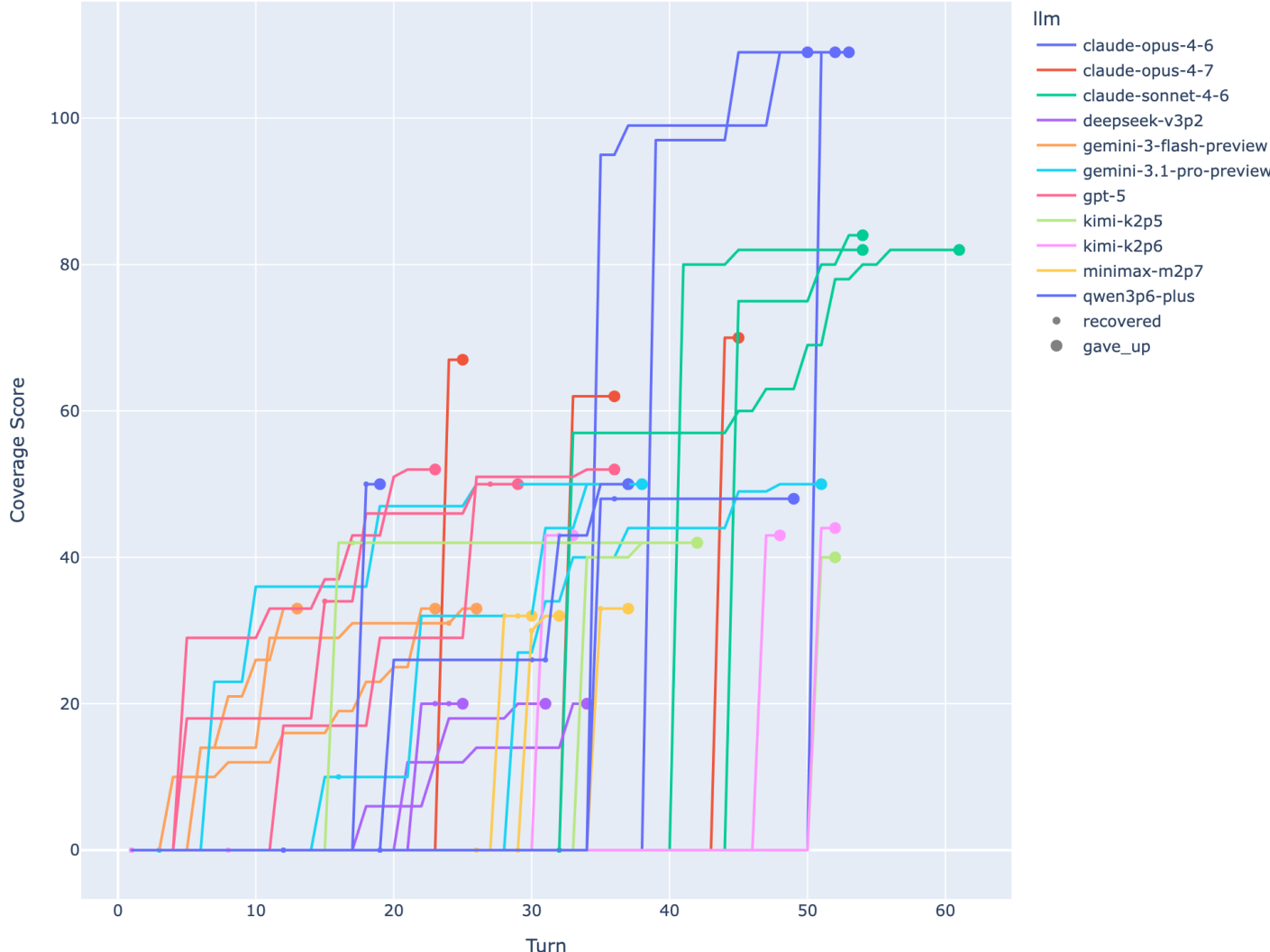

*Figure 2: Coverage Score per turn, per LLM group (sample rollouts). Claude Opus 4.6 exploration is deeper before committing; GPT-5 submits earlier. No model approaches full procedure coverage in any campaign.*

## 9.4 Tactic-Level Recall

| MITRE Tactic | Claude Opus 4.6 | Claude Sonnet 4.6 | Claude Opus 4.7 | Gemini 3.1 Pro | GPT-5 | Kimi K2.6 | Qwen3.6 Plus | Gemini 3 Flash | MiniMax M2.7 | Kimi K2.5 | DeepSeek V3.2 |
|---|---|---|---|---|---|---|---|---|---|---|---|
| Defense Evasion | 0.59 | 0.49 | 0.38 | 0.25 | 0.25 | 0.23 | 0.20 | 0.21 | 0.18 | 0.13 | 0.11 |
| Execution | 0.56 | 0.45 | 0.36 | 0.23 | 0.21 | 0.20 | 0.19 | 0.19 | 0.15 | 0.11 | 0.10 |
| Persistence | 0.56 | 0.43 | 0.34 | 0.25 | 0.27 | 0.19 | 0.21 | 0.19 | 0.16 | 0.13 | 0.14 |
| Resource Development | 0.56 | 0.44 | 0.34 | 0.26 | 0.20 | 0.20 | 0.17 | 0.21 | 0.12 | 0.14 | 0.14 |
| Command And Control | 0.55 | 0.45 | 0.41 | 0.23 | 0.23 | 0.21 | 0.22 | 0.20 | 0.16 | 0.12 | 0.11 |
| Discovery | 0.54 | 0.37 | 0.37 | 0.20 | 0.24 | 0.18 | 0.18 | 0.16 | 0.15 | 0.11 | 0.10 |
| Privilege Escalation | 0.52 | 0.41 | 0.31 | 0.22 | 0.26 | 0.20 | 0.14 | 0.18 | 0.15 | 0.11 | 0.12 |
| Initial Access | 0.37 | 0.24 | 0.19 | 0.13 | 0.02 | 0.10 | 0.12 | 0.12 | 0.04 | 0.06 | 0.04 |
| Impact | 0.33 | 0.28 | 0.24 | 0.10 | 0.22 | 0.13 | 0.15 | 0.10 | 0.11 | 0.06 | 0.07 |
| Credential Access | 0.27 | 0.20 | 0.14 | 0.09 | 0.07 | 0.12 | 0.03 | 0.06 | 0.05 | 0.06 | 0.04 |
| Lateral Movement | 0.25 | 0.18 | 0.16 | 0.08 | 0.04 | 0.09 | 0.09 | 0.06 | 0.03 | 0.05 | 0.04 |
| Collection | 0.24 | 0.16 | 0.04 | 0.09 | 0.05 | 0.07 | 0.17 | 0.06 | 0.02 | 0.04 | 0.03 |
| Exfiltration | 0.24 | 0.17 | 0.15 | 0.05 | 0.02 | 0.02 | 0.05 | 0.04 | 0.00 | 0.02 | 0.01 |

*Table 6: Normalized recall per MITRE tactic (high-severity flags only). 1.0 = maximum flags found by any model for that tactic.*

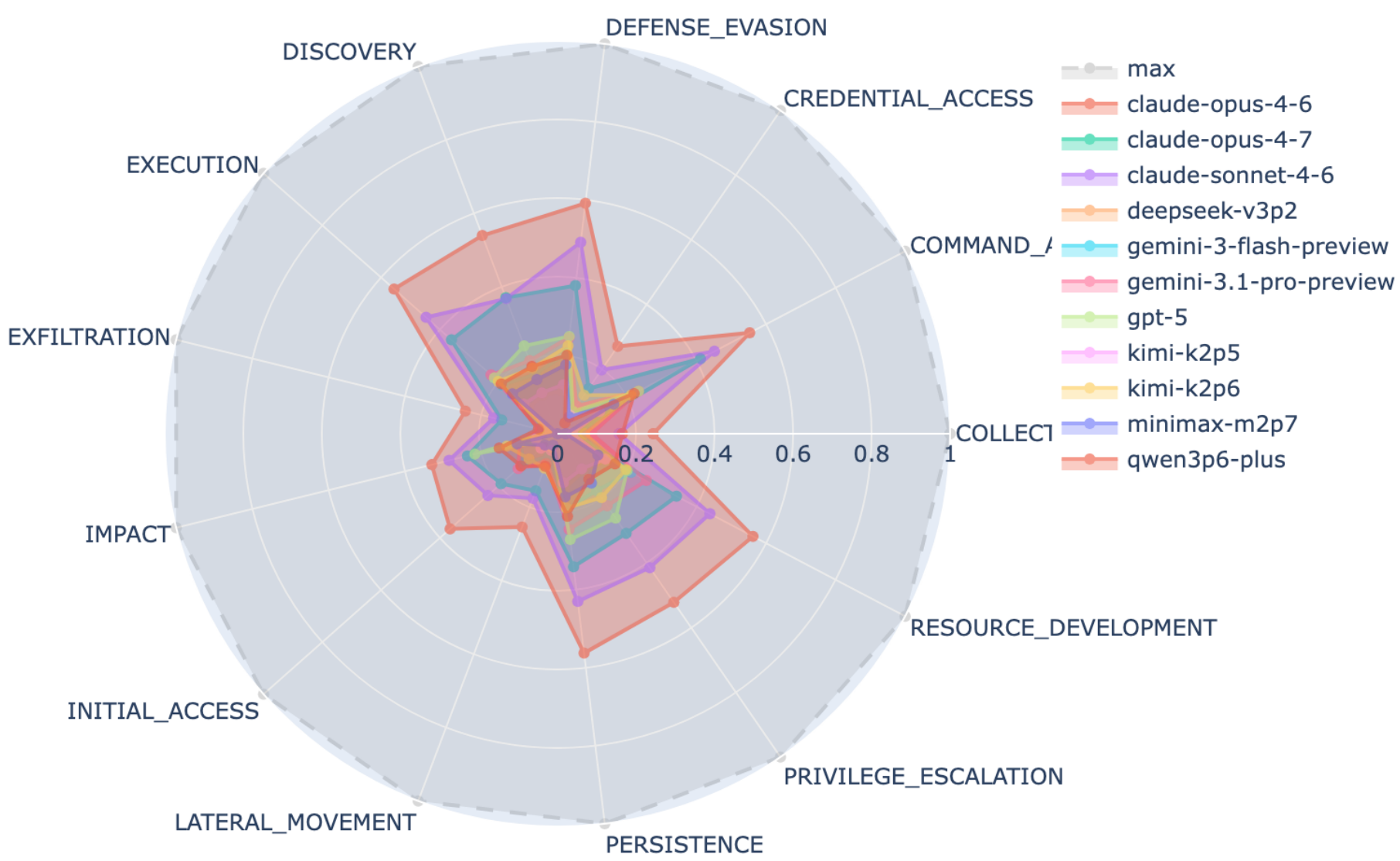

*Figure 3: Radar chart of normalized tactic recall for all five models. High-severity (critical/high), high-relevance (1-2) flags only. Claude Opus 4.6 leads across all 13 tactics; all models show near-zero coverage of Credential Access and Initial Access.*

Claude Opus 4.6 achieves its highest normalized recall on Defense Evasion (0.59) and Persistence (0.56), where common heuristics - PowerShell encoding detection, registry run-key writes, signed-binary proxy execution - are detectable with simple keyword queries. All models show near-zero recall for Credential Access and near-zero or zero recall for Initial Access, where attacks leave subtler traces (LSASS DRSUAPI calls, Kerberoasting service-ticket patterns) that require domain-specific query knowledge not spontaneously generated.

### 9.5 Hunt Strategy Analysis

Across all models, the dominant strategy is breadth-first keyword scanning: scope discovery (event counts, timestamp ranges, channel distributions), then filtering by common malicious event IDs (EventID 1 process creation, EventID 4104 PowerShell script-block logging, EventID 4624 logon events). A representative first-five-turn sequence from Claude Opus 4.6:

```
T1: SELECT COUNT(*), MIN(TimeCreated), MAX(TimeCreated) FROM logs
    -> 134,867 records, 2026-01-14T00:00:00Z to T01:51:35Z

T2: SELECT EventID, Channel, COUNT(*) AS cnt
    FROM logs GROUP BY EventID, Channel ORDER BY cnt DESC

T5: SELECT TimeCreated, EventID, Image, CommandLine FROM logs
    WHERE EventID='1' AND (
        CommandLine LIKE '%encoded%' OR
        CommandLine LIKE '%bypass%'  OR
        CommandLine LIKE '%invoke%')
    -> 16 NEW FLAGS DISCOVERED
```

GPT-5 commits to submission earlier (mean turn 34 vs. 52 for Claude Opus 4.6), suggesting a different exploration-exploitation tradeoff. Gemini 3 Flash achieves comparable flags-submitted to GPT-5 at 6× lower cost ($0.19 vs. $1.07), suggesting that reasoning quality - not token throughput or context length - is the primary bottleneck for this task.

## 10. DISCUSSION

### 10.1 Why Do Models Fail?

Three factors likely contribute to the low recall:

- **Search-space intractability.** A 10-row result limit against 75K-135K records forces agents into an information-gathering problem that cannot be solved by breadth-first scanning within 50 queries. Effective threat hunting requires semantic compression: formulating queries that retrieve high-density malicious evidence.
- **Attribution gap.** The consistent gap between n_flags_in_query and n_flags_in_submitted indicates agents observe malicious evidence but fail to attribute it. This may reflect under-confidence in partial information, aversion to submitting wrong timestamps, or limitations in tracking evidence across long conversation histories.
- **Tactic blind spots.** GPT-5 records zero recall on Initial Access and Lateral Movement; all models show near-zero recall on Credential Access. Certain ATT&CK tactics require very specific query patterns - Kerberoasting indicators in Security logs, DRSUAPI call patterns in Sysmon - that current models do not spontaneously generate without explicit detection-engineering training.

### 10.2 Benchmark Validity Considerations

The Sigma-rule-derived ground truth is comprehensive but not perfect: rules may fire on benign lookalike events, and LLM consequence enrichment introduces the evaluator's own model into the annotation pipeline. The 500-record noise injection is a conservative simulation of production log volume; real enterprise environments contain millions of daily events, further amplifying the search challenge. Microsecond-precision timestamp matching is strict by design: coarse-grained submissions (e.g., submitting all events in a one-minute window) should not receive partial credit, as the task requires precise forensic attribution.

## 11. REPRODUCIBILITY

All 26 campaign seeds, procedure labels, chain templates, simulator code, flag extraction pipeline, and evaluation harness are fully deterministic given a (seed, blueprint, timestamp) triple. The 26-seed selection procedure is implemented in research/chain_analysis/seed_simulation.py. All 859 hunt result JSONs are included in the release, containing full conversation traces, per-turn SQL queries and results, flag discovery events, token counts, and costs. Benchmark charts can be regenerated with:

```
python tools/generate_benchmark_page.py \
    --output output/benchmark.html \
    --artifacts release/artifacts/
```

LLM API results are nondeterministic at temperature > 0; single-rollout results are reported for several models. Multi-rollout averaging (≥3 rollouts) is recommended for future comparisons. The campaign generator pipeline is proprietary to Simbian AI; pre-built campaign JSONs for a benchmark seed is distributed with the release to enable independent evaluation without the generator.

## 12. LIMITATIONS AND FUTURE WORK

- **Single-rollout evaluation.** High variance suggests multi-rollout averaging is needed for statistical reliability.
- **Windows-only telemetry.** All 106 procedures use Windows event logs. Linux auditd, AWS CloudTrail, and network flow data represent a large fraction of real SOC workloads and are not covered.
- **Alert-augmented setting.** A complementary evaluation could pre-surface one Sigma-rule alert and measure the agent's ability to pivot from a known indicator of compromise, testing investigation depth separately from discovery breadth.
- **Reinforcement-learning agents.** All evaluated agents use zero-shot prompting. RL-trained agents optimizing Coverage Score may yield substantially higher recall.
- **Per-chain attribution.** The current flag set merges all malicious events regardless of sub-chain origin. Future versions will tag flags by kill-chain membership, enabling per-chain and per-stage recall metrics.

## 13. CONCLUSION

The Cyber Defense Benchmark is the first benchmark to evaluate LLM agents on open-ended, evidence-driven threat hunting against real attack telemetry. Our evaluation of five frontier models on 859 runs across 26 multi-procedure campaigns - spanning 105 of 106 MITRE ATT&CK procedures in the dataset - shows uniformly low recall: the best model (Claude Opus 4.6) reaches a Coverage Score of $0.55 \pm 0.05$ on average (the instance-weighted fraction of coverable narrative steps it surfaces per run), and no model ever completes a hunt. Against our passing bar (≥50% recall on every ATT&CK tactic), the leader clears 7 of 13 tactics and the other four models clear zero - no model passes. These results do not reflect model quality on security knowledge tasks, where the same models score well; rather, they expose a specific deficiency in open-ended, agentic evidence gathering at scale. We release the benchmark environment, campaign data, and all hunt traces to enable the community to measure and improve LLM capability in cyber defense.

## ACKNOWLEDGMENTS

We thank the Open Threat Research Federation (OTRF) for the Security-Datasets corpus, the SigmaHQ community for the open Sigma rule repository, and WithSecureLabs for the Chainsaw detection engine. This work was conducted at Simbian AI.

## APPENDIX A: BENCHMARK SEEDS

We sampled 26 seeds from the DIVERSE_INTRUSION chain generator and projected each resulting campaign into the same binary seed × procedure space used elsewhere in the report (106 procedures). From that pool we greedily picked seeds by marginal procedure gain until every reachable procedure was covered, yielding the 26-seed cohort that drives the hunt runs. This cohort reduces the rollout count by roughly a factor of 40× compared to a full-sweep evaluation and keeps API cost tractable.

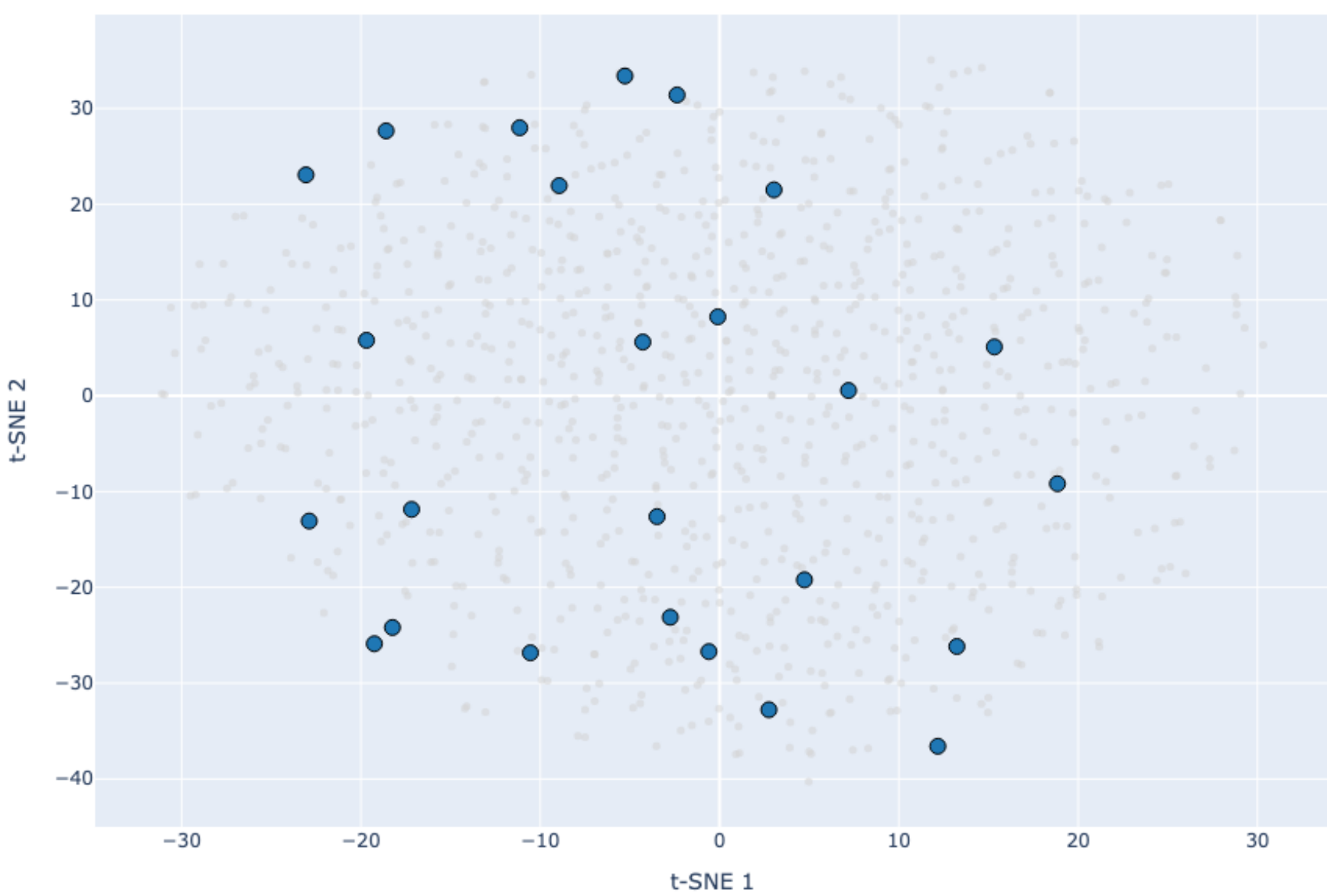

*Figure A.1: Seed universe projected to 2D via Jaccard-distance t-SNE on the binary seed × procedure one-hot matrix. Gray markers: all 1,000 seeds simulated by the generator. Blue markers: the 25-seed benchmark cohort picked via greedy marginal-gain selection. The cohort covers 105 of 106 procedures and has a mean pairwise Jaccard similarity of 0.171 (σ=0.063), within |Δ|=0.015 of the full-universe mean of 0.187 (σ=0.072) — i.e. the greedy subset stays representative of the generator distribution rather than collapsing onto a cluster.*

A sample dataset generated from one of these seeds will shipped with the GitHub release together with the code required to re-run the ReAct-based agents end-to-end, so the benchmark can be reproduced on a single seed without the full generator pipeline.

## APPENDIX B: DATASET INFRASTRUCTURE ENVIRONMENTS

| Environment | Procedure count | Primary log format(s) |
|---|---|---|
| shire_v2 | 54 | HELK (53), WEF export (1) |
| wardog_standalone | 28 | WEF export (20), raw_event (8) |
| pedro_vbox | 11 | WEF export |
| shire_v2_mordor | 4 | HELK |
| pandalab | 3 | WEF export |
| shire_v1 | 3 | Winlogbeat (2), HELK no-tags (1) |
| aptsimulator | 1 | WEF export |
| azsentinel_exchange | 1 | WEF export |
| blacksmith_adfs | 1 | WEF export |

*Table B.1: Recording infrastructure environments in the Mordor corpus.*